\documentstyle[12pt]{article}
\newcommand{\be}{\begin{equation}}
\newcommand{\ee}{\end{equation}}
\newcommand{\ba}{\begin{eqnarray}}
\newcommand{\ea}{\end{eqnarray}}
\newcommand{\baa}{\begin{eqnarray*}}
\newcommand{\eaa}{\end{eqnarray*}}
\newcommand{\lab}[1]{\label{#1}}

\newcommand{\dis}{\displaystyle}

\begin{document}
{\pagestyle{empty}
\vskip 1.5cm

{\renewcommand{\thefootnote}{\fnsymbol{footnote}}
\centerline{\large \bf Stochastic dynamics simulations in a new 
generalized ensemble}
}
\vskip 3.0cm
 
\centerline{Ulrich H.E.~Hansmann,$^{a,}$
\footnote{\ \ e-mail: hansmann@mtu.edu  Present address:
Department of Physics, Michigan Technological University,
Houghton, MI 49931, U.S.A.}
Frank Eisenmenger,$^{b,}$
\footnote{\ \ e-mail: eisenmenger@rz.hu-berlin.de}
and Yuko Okamoto$^{a,}$ \footnote{\ \ e-mail: okamotoy@ims.ac.jp}}
\vskip 1.5cm
\centerline{$^{a}$ {\it Department of 
Theoretical Studies, Institute for Molecular Science}}
\centerline{{\it Okazaki, Aichi 444-8585, Japan}}
\vskip 0.5cm
\centerline{$^b${\it Institute for Biochemistry, Medical Faculty of
the Humboldt University Berlin}}
\centerline{{\it 10115 Berlin, Germany}} 

\medbreak
\vskip 3.5cm
 
\centerline{\bf ABSTRACT}
\vskip 0.3cm

We develop a formulation for molecular dynamics,
Langevin, and hybrid Monte Carlo algorithms
in the recently proposed generalized ensemble
that is based on a physically motivated realisation of Tsallis weights.
The effectiveness of 
the methods are tested with an energy function for a protein
system.  Simulations in this generalized 
ensemble
by the three methods are performed for a penta peptide,
Met-enkephalin.  For each algorithm, it is shown that 
from only one simulation run one can
not only find the global-minimum-energy conformation 
but also obtain probability distributions in 
canonical ensemble at any temperature, which allows the calculation
of any thermodynamic quantity as a function of temperature.

\vfill
\newpage}
\baselineskip=0.8cm
\noindent
{\bf 1.~ INTRODUCTION} \\
For many important physical systems like   biological macromolecules 
it is very difficult to obtain  the accurate  canonical distribution
at low temperatures by conventional simulation methods.
This is because the energy function has many local minima,
separated by high energy barriers, and 
at low temperatures  simulations will necessarily get trapped in the
configurations corresponding to one of these local minima.
In order to overcome this multiple-minima
problem, many methods have been proposed.  
For instance, the generalized-ensemble
algorithms, most well-known of which is
the multicanonical approach, \cite{MU,MU3} 
are powerful ones and were first introduced to
the protein-folding problem in Ref.~\cite{HO}.  Simulations in the
multicanonical ensemble perform 1D random
walk in energy space.  
They can thus avoid getting trapped
in states of energy local minima.  Besides
multicanonical algorithms, simulated tempering 
\cite{ST1,ST2} and {\it 1/k}-sampling \cite{HS}
have been shown to be equally effective
generalized-ensemble methods in the protein
folding problem.\cite{HO96b}  
The simulations are usually performed with
Monte Carlo scheme, but recently molecular
dynamics version of multicanonical algorithm
was also developed.\cite{HEO96,NNK}
    
The generalized-ensemble
approach is based on non-Boltzmann probability
weight factors, and in the above three methods
the determination of the weight factors 
is non-trivial.
We have recently shown that a particular
choice of the weight factor of Tsallis statistical mechanics,\cite{Tsa}
which is a nonextensive generalization of Boltzmann-Gibbs statistical
mechanics, can be used for a 
generalized-ensemble Monte Carlo simulation.\cite{HO96d}
The advantage of this ensemble is that it greatly simplifies the 
determination of
the weight factor.

The purpose of the present work is to 
generalize this Monte Carlo approach to other simulation
techniques.
Here, we consider three commonly used algorithms:
molecular dynamics, \cite{MD} Langevin,\cite{Lang} and hybrid Monte 
Carlo.\cite{HMC}
The performances of the algorithms are tested with the system
of an oligopeptide, Met-enkephalin.\\

\noindent
{\bf 2.~ METHODS}\\
{\bf {\it 2.1.~  Monte Carlo in the new ensemble}}\\
In the canonical ensemble at temperature $T$ 
each state with potential energy $E$ is weighted by the
Boltzmann factor:
\begin{equation}
W_{B}(E,T) =  e^{-\beta E}~,
\label{eq0}
\end{equation}
where the inverse temperature $\beta$ is defined by 
$\beta = 1/k_B T$ with Boltzmann
constant $k_B$. This weight factor gives the usual bell-shaped
canonical probability distribution of energy:
\begin{equation}
P_B(E,T) \propto n(E)~W_B (E,T)~,
\label{eq1}
\end{equation}
where $n(E)$ is the density of states.
For systems with many degrees of freedom, it is usually
very difficult to generate a canonical distribution at
low temperatures.  This is because there are many local
minima in the energy function, and simulations will get
trapped in states of energy local minima.
 
Generalized-ensemble algorithms are the methods that overcome
this difficulty by performing random walks in energy space,
allowing simulations to escape from any state of energy local minimum.
Here, we discuss one of the latest examples of simulation
techniques in generalized ensemble.\cite{HO96d}
The probability weight factor of this method is given by
\begin{equation}
W (E) = \left( 1+\beta_0 \frac{E-E_{GS}}{n_F} \right)^{-n_F}~,
\label{eqopwe}
\end{equation}
where $T_0 = 1/k_B \beta_0$ is a low temperature, 
$n_F$ is the number of degrees of freedom, and
$E_{GS}$ is the global-minimum 
potential energy (when $E_{GS}$ is not known, we use its
best estimate).
Note that this weight is a
special case of the weights used in Tsallis generalized statistical mechanics,
\cite{Tsa} where the Tsallis parameter $q$ is chosen as  
\begin{equation}
q = 1 + \frac{1}{n_F}~.
\label{eqtsq}
\end{equation}
Note also that through the substraction of $E_{GS}$ it is ensured that
the weights will always be positive definite.

The above choice of $q$ was motivated by the following reasoning. 
\cite{HO96d}
We are interested in an ensemble where not only the low-energy region
can be sampled  efficiently but also the high-energy states can be visited
 with finite probability. In this way the simulation
can overcome energy barriers and escape from local minima.
The probability distribution of energy should resemble that
of an ideal low-temperature canonical distribution, but with a tail
to higher energies. The Tsallis weight at low temperature
\begin{equation}
W_{TS}(E) = {\displaystyle \left[ 1 + (q-1) \beta_0 (E - E_{GS}) 
\right]}^{- \frac{1}{q-1}}
\label{tsw2}
\end{equation}
has the required properties 
when the parameter $q$ is carefully chosen.  
Namely, 
for suitable $q > 1$ 
it is a good approximation 
of the Boltzmann weight 
$W_{B}(E,T_0) = \exp(-\beta_0 (E-E_{GS}))$
for $(q - 1) \beta_0 (E - E_{GS}) \ll 1~$, while at high energies 
it is no longer exponentially suppressed but
only according to a power law with the exponent
$1/(q - 1)$.
To ensure that simulations are able to escape from
energy local minima, the weight should start deviating from the
exponentially damped Boltzmann weight at energies near its mean value.
This is because at low temperatures there are only small fluctuations of
energy around its mean ($<E>_{T_0}$). 
In Eq.~(\ref{tsw2}) we may thus set
\begin{equation}
(q - 1)~\beta_0~(<E>_{T_0} - E_{GS}) = \frac{1}{2}~.
\label{eqnf}
\end{equation}
The mean value at low temperatures is given by the
harmonic approximation:
\begin{equation}
<E>_{T_0}  ~= E_{GS} + \frac{n_F}{2} k_B T_0 = E_{GS} + 
\frac{n_F}{2 \beta_0}~.
\label{eqhm}
\end{equation}
Substituting this value into Eq.~(\ref{eqnf}), we obtain
the optimal Tsallis parameter in Eq.~(\ref{eqtsq}).

We remark that the
calculation of the weight factor is much easier than 
in other generalized-ensemble
techniques, since it requires one to find only an estimator for the
ground-state energy $E_{GS}$, which can be done, for instance, 
by a procedure described in Ref.~\cite{HO96d}.

As in the case of other generalized ensembles,  we can
use the reweighting techniques \cite{FS} to construct canonical distributions
at various temperatures $T$.  This is because
the simulation by the present algorithm samples a large range of
energies. The thermodynamic average of any physical quantity $\cal{A}$
can be calculated over a wide temperature range by
\begin{equation}
<{\cal{A}}>_T ~=~ \frac{\displaystyle{\int dx~{\cal{A}}(x)~W^{-1}(E(x))~
                 e^{-\beta E(x)}}}
              {\displaystyle{\int dx~W^{-1}(E(x))~e^{-\beta E(x)}}}~,
\label{eqrw}
\end{equation}
where $x$ stands for configurations.\\

In the following subsections, we describe how to implement
 Langevin, molecular dynamics, and hybrid Monte Carlo algorithms in
the new ensemble defined by the weight of Eq.~(\ref{eqopwe}). 
We remark that Langevin and molecular dynamics 
algorithms for Tsallis statistical mechanics
were also developed 
in Refs.~\cite{Star} and \cite{StrMD}, respectively. 

\noindent
{\bf {\it 2.2.~ Langevin algorithm}}\\
The Langevin algorithm\cite{Lang} is used to integrate
 the following  differential equation:
\be
\dot{q}_i = -~ \dis{\frac{\partial E}{\partial q_i}} + \eta_i
 = f_i + \eta_i~, 
\lab{eq10}
\ee
where  $q_i \ (i=1,\cdots,N)$ is the (generalized) coordinates 
of the system, $E$ is the potential energy,
$f_i$ is the ``force'' acting on the particle at $q_i$,  
and $\eta_i$ is a set of independent Gaussian distributed
random variables with a variance:  
\be
< \eta_i (t_l) \eta_j (t_m) > = 2 k_B T_0 \delta_{ij} \delta(t_l - t_m).
\lab{eq10p}
\ee
Here (and hereafter), we set all the masses $m_i$ equal to unity for
simplicity.
It can be shown that 
the dynamics based on the Langevin algorithm yields a canonical
distribution $P_B (E,T_0) \propto n(E)~W_B (E,T_0)= n(E) e^{-\beta_0 E}$.  
In order to generalize this technique to simulations
in the new ensemble, we rewrite the weight factor in Eq.~(\ref{eqopwe}) as
\begin{equation}
W(E) = \exp \left\{ - \beta_0 \left[ \frac{n_F}{\beta_0} 
\ln \left(1+\beta_0 \frac{E-E_{GS}}{n_F} \right) \right] \right\},
\label{eqtswt}
\end{equation}
Defining now an effective potential energy by \cite{Star,StrMD}
\begin{equation}
 E_{eff}(E) = \frac{n_F}{\beta_0} 
\ln \left(1+\beta_0 \frac{E-E_{GS}}{n_F} \right)~,
\label{equs}
\end{equation}
we see that Langevin simulations in the new ensemble can be 
performed by replacing $E$ in Eq.~(\ref{eq10})
by $E_{eff}(E)$: 
\begin{eqnarray}
\dot{q}_i &=&  -~ \dis{\frac{\partial E_{eff}(E)}{\partial E}}
                 \dis{\frac{\partial E}{\partial q_i}} + \eta_i~,\\
&=& \frac{1}{1+\displaystyle{\displaystyle{\frac{\beta_0}{n_F}}}(E-E_{GS})}~f_i
 + \eta_i~.
\lab{eq10a}
\end{eqnarray}
Note that the procedure that led to the above equations
is exactly the same as the one we
followed when we developed molecular dynamics and related
algorithms in another generalized ensemble, i.e., multicanonical
ensemble.\cite{HEO96}

For numerical work one has to integrate the above equation by
discretizing the time with step $\Delta t$ and  therefore  for actual 
simulations we use the following difference equation:
\be
q_i(t+\Delta t) = q_i(t) + \Delta t \left(
 \frac{1}{1+\displaystyle{\displaystyle{\frac{\beta_0}{n_F}}}
\left( E(t)-E_{GS} \right)}~f_i(t)
 + \eta_i(t) \right)~. 
\lab{eq11a}
\ee
Using the above equation we will sample in the Langevin simulation
the same ensemble as in a Monte Carlo simulation with the weight of 
Eq.~(\ref{eqopwe}).
Hence, we can again use the re-weighting techniques and calculate 
thermodynamic averages according to Eq.~(\ref{eqrw}). 

\noindent
{\bf {\it 2.3.~ Molecular dynamics and hybrid Monte Carlo algorithms}}\\
Once the formulation of Langevin algorithm for the new ensemble is given,
the implementation of molecular dynamics algorithm  is 
straightforward.
    
The classical molecular dynamics algorithm is based on
a Hamiltonian 
\be
H(q,\pi) = \frac{1}{2} \sum_{i=1}^N \pi_i^2 + E(q_1,\cdots,q_N)~,
\lab{eq7}
\ee
where $\pi_i$ are the conjugate momenta corresponding to
the coordinates $q_i$. 
Hamilton's equations of motion are then given by
\be
\left\{
\begin{array}{rl}
\dot{q}_i &= \dis{\frac{\partial H}{\partial \pi_i}} = \pi_i~,\\ 
\dot{\pi}_i &= -~ \dis{\frac{\partial H}{\partial q_i}} 
           = -~ \dis{\frac{\partial E}{\partial q_i}}~=~f_i~,
\end{array}
\right.
\lab{eq8}
\ee
and they are used to generate representative ensembles of configurations.
For numerical work  the time is discretized with step $\Delta t$
and the equations are integrated according to the {\it leapfrog}
(or other time reversible integration) scheme:
\be
\left\{
\begin{array}{rl}
q_i(t+\Delta t) &= q_i(t) + \Delta t~
\pi_i \left( t + \dis{\frac{\Delta t}{2}} \right)~,\\
\pi_i \left( t+\dis{\frac{3}{2}\Delta t} \right)
&= \pi_i \left( t+\dis{\frac{\Delta t}{2}} \right) + \Delta t~
f_i(t+\Delta t)~.
\end{array}
\right.
\lab{eq12}
\ee
The initial momenta
$\{\pi_i(\frac{\Delta t}{2})\}$ for the iteration are
prepared by
\be
\pi_i \left( \dis{\frac{\Delta t}{2}} \right)
= \pi_i(0) + \dis{\frac{\Delta t}{2}} f_i (0)~,
\lab{eq13}
\ee
with appropriately chosen  $q_i(0)$ and $\pi_i (0)$
($\pi_i (0)$ is from a Gaussian distribution).

In order to generalize this widely used technique to simulations 
in our case, we again propose to replace $E$ by 
$E_{eff}$ (of Eq.~(\ref{equs})) 
 in Eq.~(\ref{eq8}). 
A new set of Hamilton's equations of motion are now given by
\be
\left\{
\begin{array}{rl}
\dot{q}_i &=~ \pi_i~, \\ 
\dot{\pi}_i &=~ -~ \dis{\frac{\partial E_{eff}}{\partial q_i}} 
 = \frac{1}{1+\displaystyle{\displaystyle{\frac{\beta_0}{n_F}}}(E-E_{GS})}~f_i~.
\end{array}
\right.
\lab{eq8a}
\ee
This is the set of equations we adopt for  MD simulation in our new
ensemble.
 For numerical work  the time is again discretized with step $\Delta t$
 and the equations are integrated according to the {\it leapfrog}
 scheme. \\

The hybrid Monte Carlo algorithm\cite{HMC} is based on the 
combination of molecular dynamics and Metropolis Monte Carlo
algorithms \cite{Metro}.  Namely, each proposal for the 
Monte Carlo update is prepared by a short MD run starting 
from the actual configuration.  In this sense, the
algorithm is based on a global update, while in the 
conventional Metropolis 
method one is usually restricted to a local update. Furthermore, 
the Metropolis
step ensures that the sampled configurations are distributed according to
the chosen ensemble, while  convential molecular dynamics simulations are
hampered by difficult-to-control systematic errors 
due to finite step size in the
integration of the equations of motion.

Given the set of coordinates $\{q_i\}$ of the previous configuration and
choosing the corresponding momenta $\{\pi_i\}$ from a Gaussian
distribution, a certain number of MD steps are performed  
to obtain a candidate configuration $\{q_i^{\prime},\pi_i^{\prime}\}$.
This candidate is accepted according to the
Metropolis Monte Carlo criterion with probability
\be
p = \min \{ 1, e^{- \beta_0 (H(q^{\prime},\pi^{\prime}) - H(q,\pi))} \}~,
\lab{eq14}
\ee
where $H$ is the Hamiltonian in Eq.~(\ref{eq7}). The time reversibility of the
{\it leapfrog} integration scheme ensures detailed balance and therefore convergence
to the correct distribution.
The whole process is repeated for a desired number of times (Monte 
Carlo steps). The number of integration ({\it leapfrog}) steps $N_{LF}$ and 
the size of the time step $\Delta t$ are free parameters in the hybrid 
Monte Carlo
algorithm, which have to be tuned carefully. A choice of 
too small $N_{LF}$ and $\Delta t$
means that the sampled configurations are too much correlated,
while too large $N_{LF}$ (or $\Delta t$) yields high rejection rates. In both cases
the algorithm becomes inefficient.
The generalization of this technique to simulations for  our ensemble
can again be made by replacing the potential energy $E$ by 
$E_{eff}$ (of Eq.~(\ref{equs})) 
in the Hamiltonian of Eq.~(\ref{eq7}).\\

\noindent
{\bf 3.~ RESULTS AND DISCUSSION}\\
The effectiveness of the algorithms presented in the previous 
section is tested for the system of an oligopeptide, Met-enkephalin.
This peptide has the amino-acid sequence Tyr-Gly-Gly-Phe-Met.
The potential energy function that we used is given 
by the sum of 
electrostatic term, Lennard-Jones term, and
hydrogen-bond term for all pairs of atoms in the peptide
together with the torsion term for all torsion angles.
The parameters for the energy function were adopted from
ECEPP/2.\cite{EC1}-\cite{EC3}  The computer code SMC \cite{SMC} was
modified to accomodate the  algorithms.  

For the generalized coordinates $\{q_i\}$ we used the dihedral angles. 
The peptide-bond dihedral angles $\omega$ were fixed to be
180$^{\circ}$ for simplicity.  This leaves 19 
dihedral angles as generalized coordinates ($n_F = 19$).
The global-minimum potential energy $E_{GS}$ in this case was
obtained previously and we have $E_{GS} = -10.7$ kcal/mol.\cite{EH96d}
As for the temperature, we set 
$T_0 = 50$ K (or, $\beta_0 = 10.1$ $[\frac{1}{{\rm kcal}/{\rm mol}}]$),
following the case for the Monte Carlo simulation in the
present generalized ensemble.\cite{HO96d}
We used these numerical values for $n_F$, $E_{GS}$, and $\beta_0$
in Eq.~(\ref{eqopwe}).
We remark that the convention for energy values in the present code, SMC,
\cite{SMC} is slightly different from the one in the previous
Monte Carlo work.\cite{HO96d}  Thus, the above value of $E_{GS}$ is
accordingly different from that in Ref.~\cite{HO96d}. 
By the definition of generalized ensembles, which are based on
non-Boltzmann weight factors, one cannot obtain information
on the real dynamics of the system by the MD algorithm, and
only static thermodynamic quantities can be calculated.
For this reason we do not need to consider the 
equations of motion for dihedral space
as presented in Ref.~\cite{MDA}, but can use the much 
simpler form for the kinetic energy term as given in
the previous section (see Eq.~(\ref{eq7})). 

For the  MD simulations in our ensemble, we made a single
production run with the total number of time 
steps $N_{LF} = 800,000 \times 19$ and the 
time-step size $\Delta t = 0.0075$ (in arbitrary units). 
For  Langevin algorithm,
a production run with the same number of time steps 
($N_{LF} = 800,000 \times 19$) as in
the MD simulation was performed, but our optimal  
time-step size was only $\Delta t = 0.00028$.  This 
indicates that the simulation moves more slowly
through phase space, and we expect slower convergence
to the  Tsallis distribution than in MD case.
For the 
hybrid Monte Carlo algorithm, an MD simulation with 
19 leapfrog steps was made for each Monte Carlo step and
a production run with 400,000 MC steps was made. Since the 
Metropolis step 
in hybrid Monte Carlo corrects for errors due to 
the numerical integration
of the equation of motion, the time-step size 
$\Delta t$ can be large for 
this algorithm. We
chose $\Delta t = 0.01375$ in our units. The initial 
conformation
for all three simulations was the final (and therefore equilibrized) 
conformation obtained from
a Monte Carlo simulation of 200,000 sweeps, 
following 1,000 sweeps for thermalization with the same weight (in
each sweep all of the 19 angles were updated once). 

In Fig.~1 the time series of the total potential energy are
shown for the three  simulations with the new weight.  They all
display a random walk in energy space as they should for a simulation
with the weight of Eq.~(\ref{eqopwe}).  All the lowest-energy conformations 
obtained were essentially the same (with only a small
amount of deviations for each dihedral angle) as that of the 
global-minimum
energy conformation previously obtained for the same
energy function (with $\omega = 180^{\circ}$) by other 
methods.\cite{OKK,HO,MMV}  The global-minimum potential energy value 
obtained by
minimization is $-10.7$ kcal/mol.\cite{MMV} 
The random walks of the MD and hybrid MC simulations visited
the global-minimum region ($E < -10$ kcal/mol) six times 
and eight times, respectively, while
that of the Langevin simulation reached the region only twice.
These visits are separated by the walks towards the high-energy region
much above $E = 16$ kcal/mol, which corresponds to the average energy 
at $T=1000$ K.\cite{HO} 
Hence, the rate of convergence to the generalized ensemble
is the same order for all three methods (with MD and Langevin
algorithms slightly slower).  As discussed below, however, the
results of thermodynamic quantity calculations all agree with
each other, implying that the methods are equally reliable.

In Fig.~2 the time series of the overlap $O$ of the
conformation with the
ground state is plotted.
Our definition of the overlap, which measures how similar a given 
conformation is to the lowest-energy conformation, is given by
\begin{equation}
O(t) = 1 -\frac{1}{90~n_F} \sum_{i=1}^{n_F} 
\left| \theta_i^{(t)}- \theta_i^{(GS)} \right|~,
\label{eqol}
\end{equation}
where $\theta_i^{(t)}$ and $\theta_i^{(GS)}$ (in degrees) stand for 
the $n_F$ torsion angles of the conformation at $t$-th simulation step
and the lowest-energy conformation, respectively.  Symmetries
for the side-chain angles were taken into account and the difference
$\theta_i^{(t)}- \theta_i^{(GS)}$ was always projected into the interval
$[-180^{\circ},180^{\circ}]$. Our definition  guarantees that we have 
\begin{equation}
0 \le ~<O>_T~ \le 1~,
\end{equation}
with the limiting values
\begin{equation}
\left\{
\begin{array}{rl}
 <O(t)>_T~~ \rightarrow 1~,~~&T \rightarrow 0~, \\
 <O(t)>_T~~ \rightarrow 0~,~~&T \rightarrow \infty~.
\end{array}
\right.
\label{eqordp}
\end{equation}
Only the result from the  hybrid Monte Carlo simulation
is given in Fig.~2, since the other two simulations give similar results.
Note that there is a clear anti-correlation between the
potential energy $E$
and overlap $O$ (compare Figs.~1c and 2), indicating that 
the lower the potential energy is, the larger the overlap is
(closer to the ground state).

Simulations in generalized ensemble can not only find the energy
global minimum but also any thermodynamic quantity as a function
of temperature from a single simulation run.  As an example,
we show in Fig.~3
the average potential energy as a function of
temperature calculated from independent runs of the 
three  algorithms together with that 
from  Monte Carlo results of Ref.~\cite{EH96d} which rely on
a multicanonical Monte Carlo simulation. The results all agree 
within error bars.
      
Another example of such a calculation is the average overlap
as a function of temperature.  The results are essentially the
same for the three  algorithms.  That from the MD algorithm 
is shown in Fig.~4.  
We see that the average overlap approaches 1
in the limit the temperature going to zero, as it should
(see Eq.~(\ref{eqordp})).
We remark that the average overlap
approaches the other limiting
value, zero (see Eq.~(\ref{eqordp})),
only very slowly as the temperature increases. This is 
because $<O>_T ~= 0$ corresponds to
a completely random distribution of dihedral angles which is energetically 
highly unfavorable because of the steric hindrance of both main and side
chains.\\ 

\noindent
{\bf CONCLUSIONS} \\
In this article we have  shown that the generalized-ensemble
algorithm based on a special realisation of Tsallis weights is
not restricted to Monte Carlo simulations, but can also be used in 
combination with other simulation methods such as molecular dynamics,
Langevin, and hybrid Monte Carlo algorithms.
We have tested the performances of the above three methods in 
the generalized 
ensemble for a simple peptide, Met-enkephalin.  The results were
comparable to those of the original Monte Carlo version \cite{HO96d}  
in that the rate of convergence to the generalized ensemble is
of the same order and that the thermodynamic quantities calculated as
functions of temperature all agree with each other.
We believe that there is a wide range of applications for 
the generalized-ensemble
versions of molecular dynamics and related algorithms.  
For instance, the generalized-ensemble MD simulations may 
prove to be a valuable tool 
 for refinement of protein structures
inferred from X-ray and/or NMR experiments.  

\vspace{0.5cm}
\noindent
{\bf Acknowledgements}: \\
Our simulations were performed on the computers of the Computer
Center at the Institute for Molecular Science, Okazaki,
Japan.  This work is
supported, in part, by a Grant-in-Aid for Scientific Research
from the
Japanese Ministry of Education, Science, Sports and Culture.


\newpage
\noindent
\centerline{\bf Figure Captions}

\begin{itemize}
\item FIG.~1. (a) Time series of the total potential energy $E$ (kcal/mol)
from a Langevin simulation 
in the new generalized ensemble.  
The simulation consists of 
$800,000 \times 19$ time steps with step size $\Delta t = 0.000028 $.
(b) Time series of $E$ from a molecular dynamics simulation  
in the new generalized ensemble.  
The simulation consists of 
$800,000 \times 19$ time steps with step size $\Delta t = 0.0075 $.
(c) Time series of $E$ from a hybrid Monte Carlo simulation 
in the new generalized ensemble.  
The simulation consists of 
        400,000 MC steps.  For each MC step an MD run of 19 
time steps was made with step size $\Delta t = 0.01375$.
\item FIG.~2.  Time series of overlap function  $O$ (as defined in the text)
        from the hybrid Monte Carlo simulation
        in the new generalized ensemble.
The simulation consists of 
        400,000 MC steps.  For each MC step an MD run of 19 
time steps was made with step size $\Delta t = 0.01375$.
\item Fig.~3.  The average potential energy $<E>_T$ (kcal/mol) 
as a function of temperature $T$ (K)
obtained from independent runs of the three algorithms and  a multicanonical  
 Monte Carlo simulation of Ref.~\cite{EH96d}
\item FIG.~4:  The average overlap function $<O>_T$ (defined in the text) as a 
function of 
temperature $T$ (K) obtained
from the  molecular dynamics simulation in the new generalized
ensemble.
\end{itemize}
   
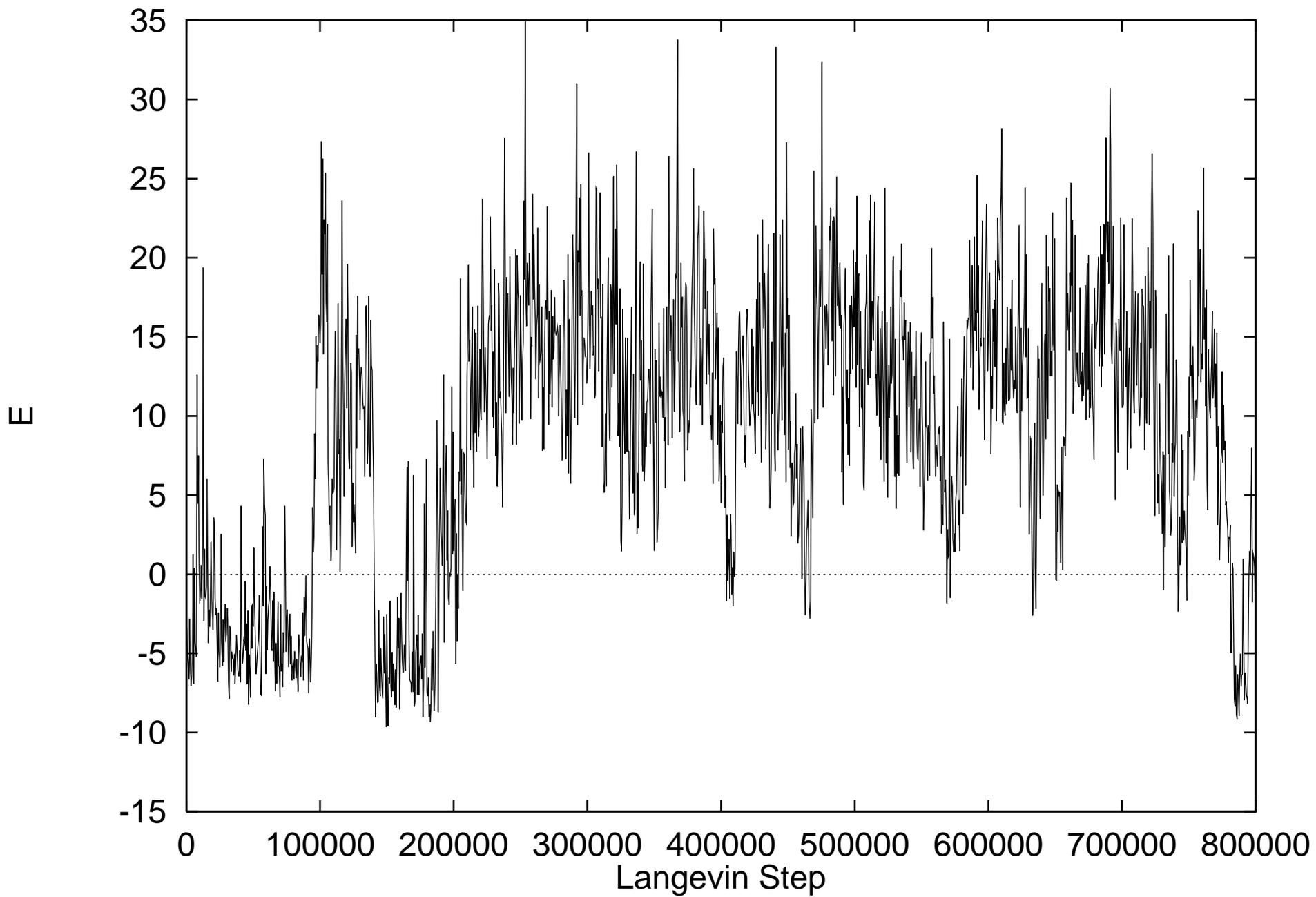
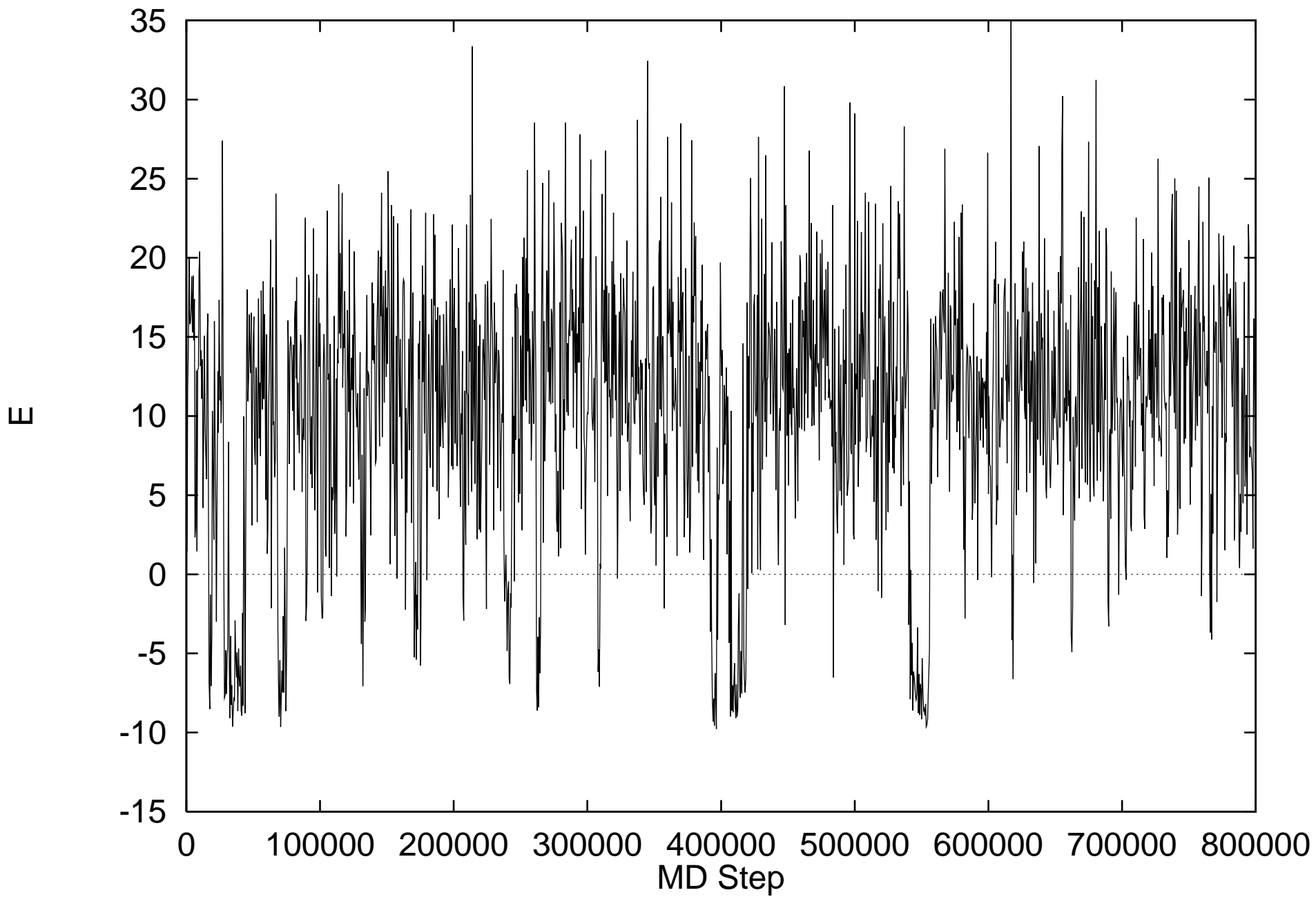
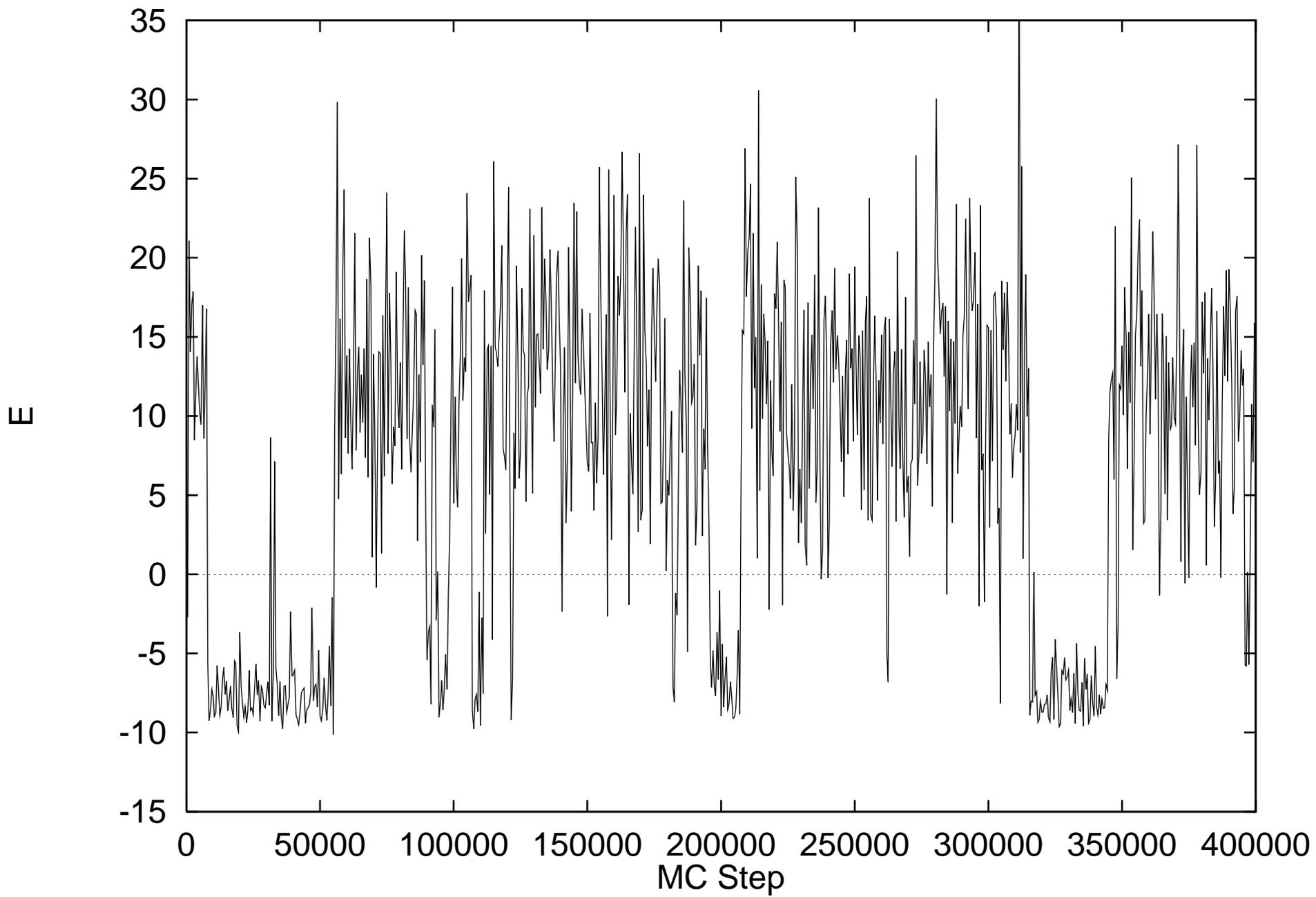
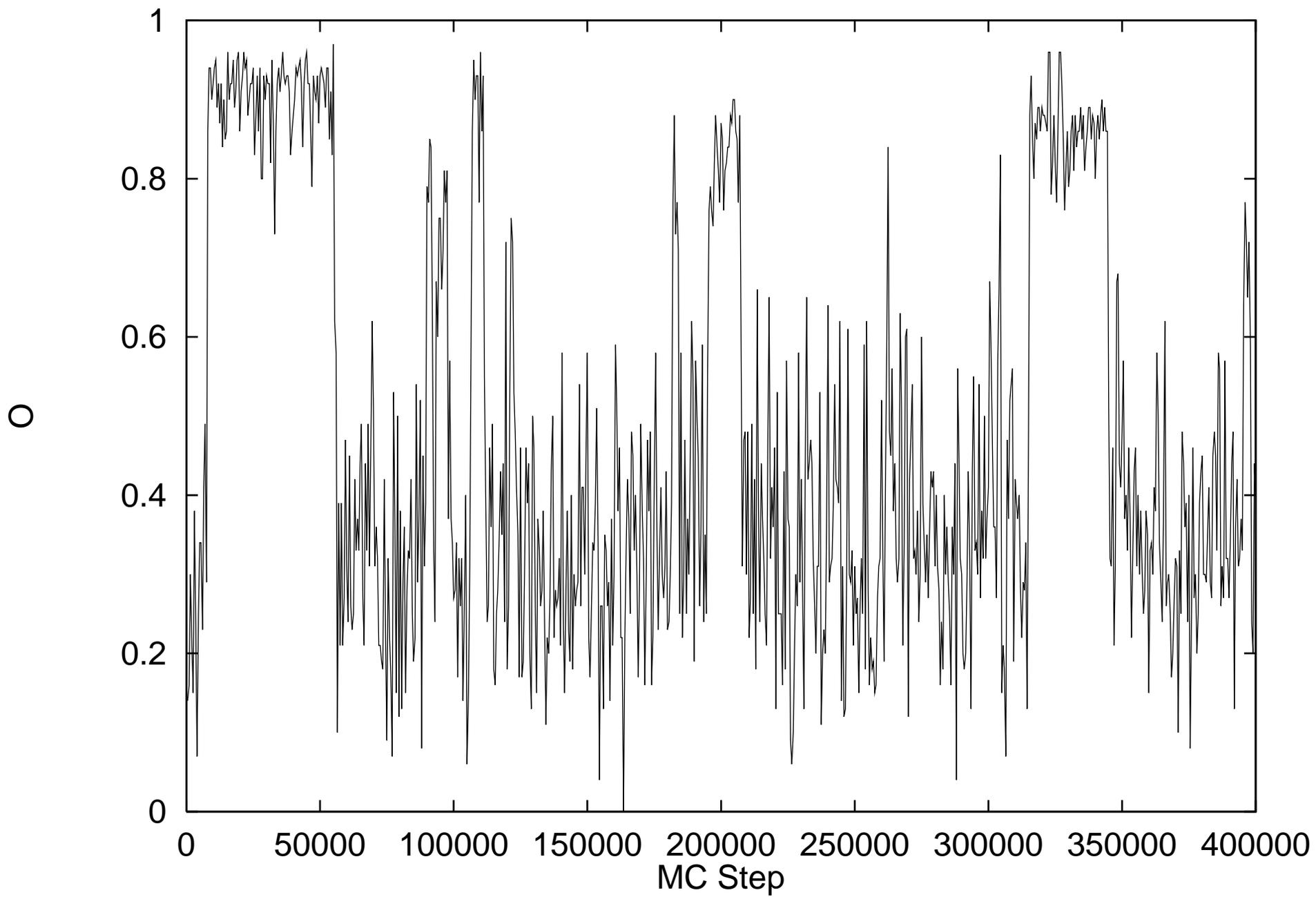
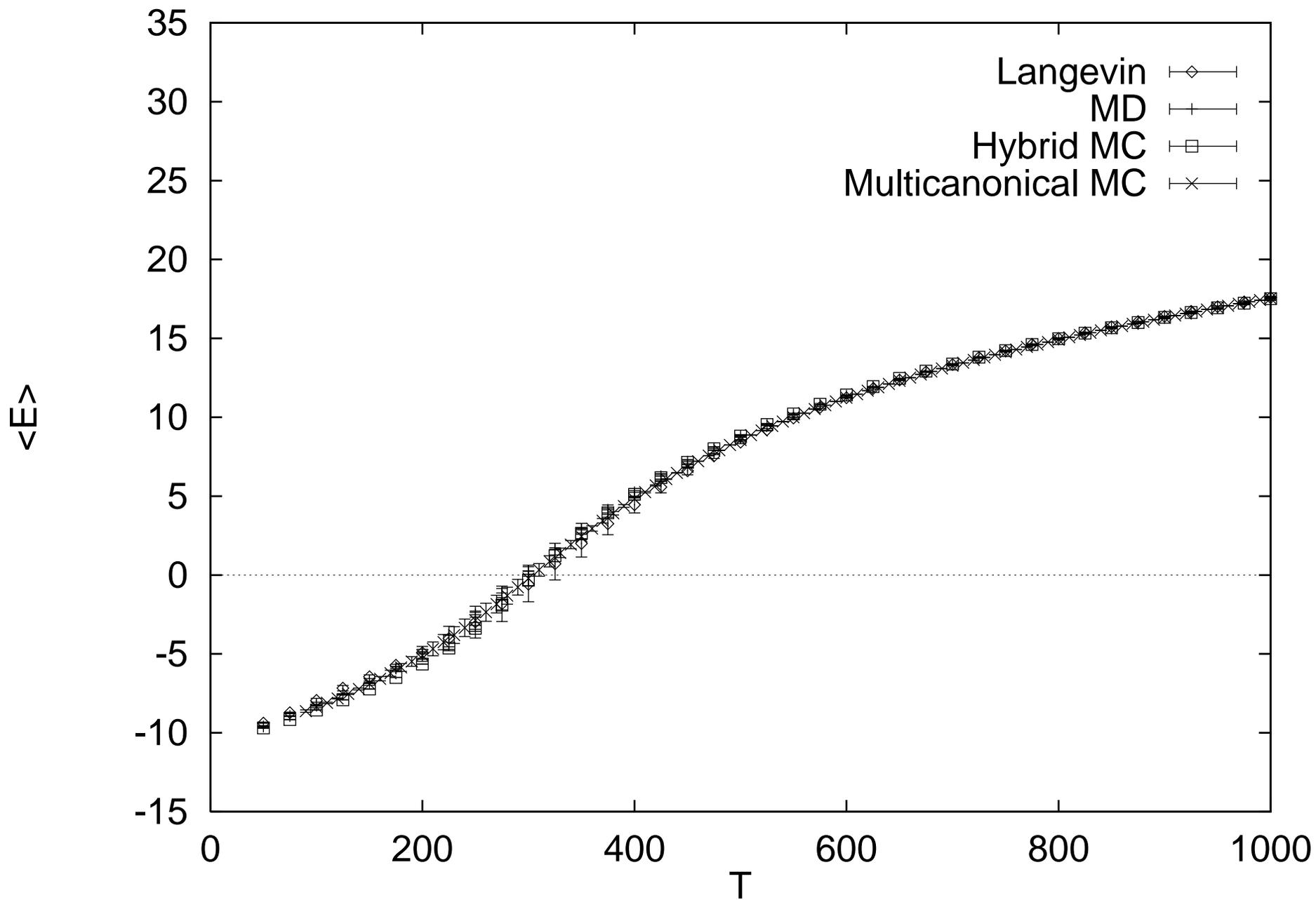
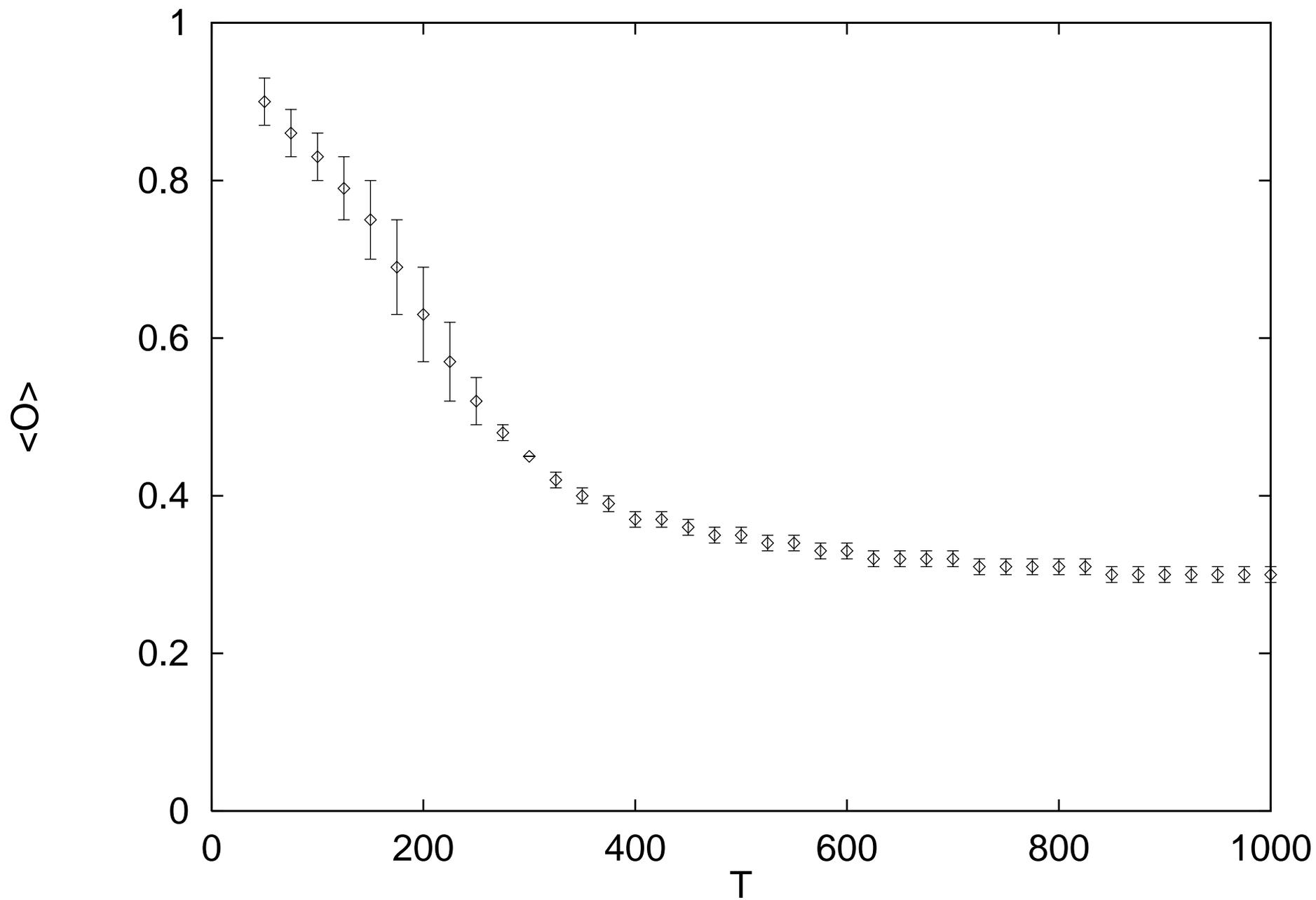
\end{document}